\documentstyle[12pt]{article}
\newcommand{\sls}[1]{/\!\!\!#1}
\newcommand{\be}{\begin{equation}}
\newcommand{\ee}{\end{equation}}
\newcommand{\ba}{\begin{eqnarray}}
\newcommand{\ea}{\end{eqnarray}}

\begin{document}

\begin{titlepage}
\leftline{\_\hrulefill\kern-.5em\_}
\vskip -4 truemm
\leftline{\_\hrulefill\kern-.5em\_}
\centerline{\small{May 1992\hfill
Dipartimento di Fisica dell'Universit\`a di Pisa\hfill
IFUP-TH \ 20/92}}
\vskip -2.7 truemm
\leftline{\_\hrulefill\kern-.5em\_}
\vskip -4 truemm
\leftline{\_\hrulefill\kern-.5em\_}
\vskip1.5truecm
\centerline{Radiative correction effects of a very heavy top}
\vskip.7truecm
\centerline{
${\rm Riccardo\:Barbieri}^{\,1,\,2,\,\dagger}$ \,
${\rm Matteo\:Beccaria}^{\,1,\,2}$ \,
${\rm Paolo\:Ciafaloni}^{\,1,\,2}$}
\centerline{
${\rm Giuseppe\: Curci}^{\,2,\,1}$ \,
${\rm Andrea\:Vicer\acute{e}}^{\,3,\,2}$}
\vskip1truecm
\centerline{\footnotesize{(1) Dipartimento di Fisica, Universit\'a di Pisa}}
\centerline{\footnotesize{Piazza Torricelli 2, I-56100 Pisa, Italy}}
\vskip1truemm
\centerline{\footnotesize{(2) I.N.F.N., sez. di Pisa}}
\centerline{\footnotesize{Via Livornese 582/a, I-56010 S. Piero a
Grado (Pisa) Italy}}
\vskip1truemm
\centerline{\footnotesize{(3) Scuola Normale Superiore}}
\centerline{\footnotesize{Piazza dei Cavalieri 7, I-56100 Pisa, Italy}}
\vskip1truecm
\begin{abstract}
\small{
If the top is very heavy, $m_t\gg M_Z$, the dominant radiative correction
effects in all electroweak precision tests can be exactly characterized in
terms of two quantities, the $\rho$-parameter and the GIM violating $Z\to
b\bar{b}$ coupling. These quantities can be computed using the Standard
Model Lagrangian with vanishing gauge couplings. This is done here up to
two loops for arbitrary values of the Higgs mass.
}
\end{abstract}
\vfill
$\dagger$ Address from May 1, 1992: CERN, Geneva, Switzerland
\end{titlepage}
%
%\section{ }
%
{\bf 1.\,}
The effects of virtual heavy top quark exchanges in electroweak precision
tests are recognized as being very important. In the Standard Model of the
electroweak interactions these effects are used to get a significant
constraint on the top quark mass, $m_t$. Perhaps even more importantly,
these effects are so large that, until the value of $m_t$ will not be
measured directly and independently, they will obscure the comparison of
different models, including the SM itself, in their predictions of
electroweak precision tests. We have especially in mind the effects that
grow like $m_t^2$ at one loop level, which affect the electroweak
$\rho-$parameter~\cite{1} ( and all related quantities ) and the
GIM-violating coupling of the $Z$ boson to a $b\bar{b}$ pair~\cite{2}. The
main contribution of this work is the explicit calculation of the two loop
$m_t^4-$corrections to these quantities in the SM for arbitrary values of
the Higgs mass, $m_H$. In the literature~\cite{3} one finds already the
$m_t^4$ contributions to $\rho$ for $m_H \ll m_t$, which we confirm.

The actual calculation of these corrections is greatly simplified by the
observation that to obtain them it is enough to consider the lagrangian of
the SM in the limit of vanishing coupling constants: the gauge bosons play
the r\^ole of external sources and the relevant propagating fields are the
top quark, the massless bottom quark, the Higgs field and the charged and
neutral Goldstone bosons, $\phi^\pm, \chi$. We call this the Gaugeless
Limit of the Standard Model.

\vskip .5 truecm
{\bf 2.\,}
A way to relate proper quantities computed in the reduced model to physical
observables is to consider the Ward Identities satisfied by the charged
weak current $J_\mu^\pm$ and by the usual combination of neutral currents,
$J_\mu = J_\mu^3 - \sin^2\theta_W J_\mu^{em}$,
that couples to the $Z$ boson.
When the bare neutral component of the
Higgs doublet acquires a vev $v$, both these currents get a piece
proportional to the derivative of the Goldstone fields
\ba
J_\mu &=& \hat{J}_\mu +\, \frac{v}{\sqrt{2}}\,\, \partial_\mu \chi \\
J_\mu^\pm &=& \hat{J}^\pm_\mu \mp\, i\,  v\,\,
\partial_\mu
\varphi^\pm \nonumber
\ea
{}From the current conservation eq.s $\partial_\mu J_\mu = \partial_\mu
J_\mu^\pm=0$, it is easy to derive the following Ward Identities
($s^2 \equiv \sin^2\theta_W$, $P_{L, R} = \frac{1}{2}(1\pm\gamma_5)$)
\ba
(p-p^\prime)_\mu\, \Gamma_\mu(p,\,p^\prime) &=&
\frac{iv}{\sqrt{2}} \Gamma(p,\, p^\prime)  +   \nonumber \\
&&
+\frac{1}{2}\left[\left(1-\frac{2}{3}s^2\right)
\left(
S^{-1}(p^\prime)\,\, P_L-
P_R\,\, S^{-1}(p)\right)+ \right. \nonumber \\
&&
\left. -\frac{2}{3}s^2
\left(
S^{-1}(p^\prime)\,\, P_R-
P_L\,\, S^{-1}(p)\right)\right] \nonumber \\
q_\mu q_\nu\, \Pi_{\mu\nu}(q) &=& \frac{v^2}{2}\, \Pi(q)
\label{WI}\\
q_\mu q_\nu\, \Pi^\pm_{\mu\nu}(q) &=& v^2\, \Pi^\pm(q)
\nonumber
\ea
which involve the $b$-quark propagator $S(p)$;
the self energies, $\Pi(q)$, $\Pi^\pm(q)$, of the
neutral and charged Goldstones; the correlation functions
$\Pi_{\mu\nu}(q)$, $\Pi_{\mu\nu}^\pm(q)$ of the currents $\hat{J}_\mu$,
$\hat{J}_\mu^\pm$ respectively;
the vertex $\Gamma_\mu(p,\,p^\prime)$
between $b$-quarks of momentum $p$, $p^\prime$ and
the current $\hat{J}_\mu$; the
vertex $\Gamma(p, \, p^\prime)$
between $b$-quarks of momentum $p$, $p^\prime$ and the neutral
Goldstone $\chi$.
Both the $\Pi$'s and the $\Gamma$'s are meant to be irreducible.

To get the physical observables we now define the following constants
\ba
\Gamma_\mu(p,\, p^\prime) &\simeq& -\frac{i}{2}
\, \left[
\left(1-\frac{2}{3} s^2\right)\, Z_1\, \gamma_\mu P_L
-\frac{2}{3} s^2\gamma_\mu P_R\right]
\ \
\mbox{at}\ \ p\simeq p^\prime
\nonumber \\ \nonumber \\
\Gamma(p,\, p^\prime) &\simeq& Z_1^\chi\,
\frac{\sls{p}^\prime-\sls{p}}
{\sqrt{2}\, v}\, P_L \ \
\mbox{at}\ \ p\simeq p^\prime
\nonumber \\ \nonumber \\
S^{-1}_b(p) &\simeq& i\,Z_2^b\, \sls{p} P_L
+i\sls{p} P_R
\ \
\mbox{at}\ \ p^2\simeq 0
\nonumber \\ \nonumber \\
\Pi_{\mu\nu}(q) &\simeq& \frac{v^2}{2}\, (Z-1)\, \delta_{\mu\nu} \ \
\mbox{at}\ \ q\simeq 0
\\ \nonumber \\
\Pi_{\mu\nu}^\pm(q) &\simeq& v^2\, (Z^\pm-1)\,
\delta_{\mu\nu} \ \
\mbox{at}\ \ q\simeq 0
\nonumber \\ \nonumber \\
\Pi(q) &\simeq& (Z_2^\chi-1)\, q^2 \ \
\mbox{at}\ \ q^2\simeq 0
\nonumber \\ \nonumber \\
\Pi^\pm(q) &\simeq& (Z_2^\varphi-1)\, q^2 \ \
\mbox{at}\ \ q^2\simeq 0
\nonumber
\ea
They satisfy, from the Ward Identities (\ref{WI}), the relations
\be
\left(1-\frac{2}{3} s^2\right)\,Z_1 = Z_1^\chi +
\left(1-\frac{2}{3} s^2\right)\,Z_2^b\,\,; \ \ \ \
Z = Z_2^\chi\,\,; \ \ \ \
Z^\pm = Z_2^\varphi
\ee
By recalling now that, in the full SM Lagrangian, the $W$- and $Z$-bosons
couple to the current $J_\mu^\pm$, $J_\mu$ via
\be
\Delta {\cal L} = \frac{g}{\sqrt{2}}\left(J_\mu^+ W_\mu^- +
J_\mu^- W_\mu^+\right) + \frac{g}{c} J_\mu Z_\mu
\ee
these relations enable us to compute, to leading order in the $SU(2)$ gauge
coupling $g$, the physical $Z_\mu\to b\bar{b}$ vertex
\be
\label{vertex}
V_\mu = -i\,\frac{g}{2\,c}
\, \left[
\left(1-\frac{2}{3} s^2 + \frac{Z^\chi_1}{Z_2^b}\right)
\,\gamma_\mu P_L -\frac{2}{3} s^2 \gamma_\mu P_R\right]
\ee
and the vector boson masses
\ba
\label{masses1}
M_Z^2 &=& \frac{g^2 v^2}{2\,c^2}\, Z = \frac{g^2 v^2}{2\, c^2}\, Z_2^\chi
\nonumber \\
\\
\label{masses2}
M_W^2 &=& \frac{g^2 v^2}{2}\, Z^\pm = \frac{g^2 v^2}{2}\, Z_2^\varphi
\nonumber \\
\ea
These are indeed the physical masses, as $g\to 0$, because the
displacement between $q^2 = 0$ and the pole at $q^2 = M^2$ is irrelevant
and because there is no wave function renormalization of the vector bosons
in this limit. The $\rho$ parameter is therefore \cite{4}
\be
\label{rho}
\rho = \frac{M_W^2}{M_Z^2\, c^2} = \frac{Z_2^\varphi}{Z_2^\chi}
\ee
Notice that the Ward Identities (\ref{WI}) are identically true, as they
stand, also in the full theory, with the gauge couplings switched on,
if one works in the Background Gauge.
Let us also remark that, if we had used an effective Lagrangian formalism,
the constants $Z_1^\chi$, $Z_2^\chi$, $Z_2^\varphi$ would have appeared in
front of terms involving derivatives of the Goldstone bosons. Eq.s
(\ref{vertex}-\ref{masses2}) could have then be simply obtained by proper
covariantization of these derivatives. Since the $W$ and the $Z$ are
treated as external sources, there is never the need to fix the gauge and
break gauge invariance.

What remains to be done at this point is the re-expression of the various
parameters appearing in (\ref{vertex}) and (\ref{rho}) in favor of
physically measurable quantities. As usual, $g$, $v$ and $c$ are traded for
$M_Z$, eq. (\ref{masses1}-\ref{masses2}), the
fine structure constant $\alpha$ and the
Fermi constant $G_\mu$ as measured in $\mu$-decay, which, in our
approximation, are given by
\be
\label{alphaandG}
\alpha = \frac{g^2}{4\pi s^2}, \ \ \ \
G_\mu = \frac{\sqrt{2} g^2}{8 M_W^2} = \frac{\sqrt{2}}{4 v^2 Z_2^\varphi}
\ee
{}From eq.s (\ref{vertex}-\ref{alphaandG}), by
retaining only the non vanishing corrections as $g\to 0$, the radiative
effects in all electroweak precision tests can be ``non-perturbatively''
characterized in terms of the two quantities $\rho$ and $\tau\equiv
Z_1^\chi/Z_2^b$ appearing in the GIM-violating $Z\to b\bar{b}$ vertex.
In particular, wherever $s^2$ appears, it must be replaced, from eq.s
(\ref{masses1}-\ref{alphaandG}), with
\be
s^2 = \frac{1}{2}\left(\,\,1-\sqrt{1-\frac{4\pi\alpha}{\sqrt{2} G_\mu
M_Z^2\rho}\,\,}\,\,\right)
\ee
As an example, in terms of $\rho$ and $\tau$, the width of the $Z$ into a
$b\bar{b}$ pair is given by
\ba
\Gamma\left(Z\to b\bar{b}\right) &=& \rho\frac{G_\mu M_Z^3}{8\pi\sqrt{2}}
\sqrt{1-\frac{4m_b^2}{M_Z^2}}\left[\left(g_{bV}^2 + g_{bA}^2\right)
\left(1+2\frac{m_b^2}{M_Z^2}\right)-6g_{bA}^2\frac{m_b^2}{M_Z^2}\right]
\nonumber \\
g_{bV} &=& 1-\frac{4}{3}s^2 + \tau, \ \ \ \ \
g_{bA} = 1 + \tau
\ea
where the kinematical dependence on the $b$ quark mass, $m_b$, is
also introduced.
Of course, hidden
in $\rho$ and $\tau$ there are the top Yukawa coupling $g_t$
and the quartic Higgs coupling $\lambda$, which must also be expressed in
favor of the top and of the Higgs mass, defined as the positions of the
poles of the corresponding propagators. From the top propagator
\be
S_t^{-1}(p) \simeq i\, Z_{2L}^t\, \sls{p}\, P_L +
i\, Z_{2R}^t\, \sls{p}\,
P_R + g_t\, v\, B\ \ \ \  \mbox{at}\ \ \sls{p}\simeq m_t
\ee
one has
\be
\label{redefinition}
m_t = g_t v B/(Z_{2L}^t Z_{2R}^t)^{1/2}
\ee
whereas, up to the order
we are working, $m_H^2 = 4\lambda v^2$

\vskip .5 truecm
{\bf 3.\,}
Fig 1 contains the relevant one loop diagrams contributing to
the constants $Z_2^\chi$, $Z_2^\varphi$ and $Z_1^\chi$ which
allow to determine the coefficients of the first term in an expansion
in powers of $G_\mu m_t^2$ both of $\rho$ and of the GIM violating
$Z\to b\bar{b}$ vertex. The diagrams contributing to $Z_2^\chi$ and
$Z_2^\varphi$ and only containing Higgs or Goldstone boson lines, but no
quark lines, have not been drawn since they do not affect $\rho$ or more
precisely, the difference $Z_2^\chi - Z_2^\varphi$, in terms of which
$\rho$
can be identically written, using eq.s (\ref{rho}, \ref{alphaandG}), as
\be
\frac{1}{\rho}-1 = \frac{4 G_\mu v^2}{\sqrt{2}}\left(Z_2^\chi - Z_2^\varphi
\right)
\ee
In the literature \cite{5}, this is often called the irreducible part of
the corrections to the $\rho$ parameter.
At order $\left(G_\mu m_t^2\right)^2$, other than the contributions of the
genuine two loops irreducible diagrams, one has to compute the $G_\mu
m_t^2$ corrections to the wave function renormalizations of the top and
bottom quarks as specified in eq.s (\ref{vertex}, \ref{redefinition}).

In a decently compact way, the results of the calculations can be given in
an analytic form in two different
asymptotic regimes ( $ x = \frac{G_\mu m_t^2}{8 \pi^2\sqrt{2}}$, $N_c = 3$,
$ r = \frac{m_t^2}{m_H^2}$ )

\begin{description}

\item[(a) $m_t \gg m_H$]

\begin{eqnarray}
\frac{1}{\rho} - 1 &=& -N_c x \left[1 + x \left(19 - 2
\pi^2\right)\right]\nonumber\\
\frac{Z_1^\chi}{Z_2^b} &=& -2 x \left[1 +
\frac{x}{3}\left(27 - \pi^2\right)\right]
\label{9}
\end{eqnarray}

\item[(b) $m_H \gg m_t$]
\begin{eqnarray}
\lefteqn{\frac{1}{\rho} - 1 =} && \ \ \nonumber\\
&& -N_c x \left\{1 + x \left[ \frac{49}{4} + \pi^2 + \frac{27}{2}\log{r} +
\frac{3}{2}\log^2{r} +\right.\right.\nonumber\\
&& \qquad\qquad + \frac{r}{3} \left(2 - 12 \pi^2 + 12 \log{r} - 27
\log^2{r}\right) + \nonumber\\
&&\qquad\qquad\left.\left. +\frac{r^2}{48}\left(1613 - 240 \pi^2 - 1500
\log{r} - 720 \log^2{r}\right)\right]\right\}\nonumber\\
\lefteqn{\frac{Z_1^\chi}{Z_2^b} =} && \ \ \nonumber\\
&& -2 x \left\{ 1 + \frac{x}{144} \left[311 + 4 \pi^2 + 282 \log{r} + 90
\log^2{r} + \right.\right.\nonumber\\
&&\qquad\qquad - 4 r \left(40 + 6 \pi^2 + 15 \log{r} + 18\log^2{r}\right)
+\nonumber\\
&&\left.\left.\qquad\qquad + \frac{3 r^2}{100} \left(24209 - 6000 \pi^2 -
45420 \log{r} - 18000 \log^2{r}\right)\right]\right\}\nonumber\\
\label{10}
\end{eqnarray}

\end{description}

In the expression for $m_H \gg m_t$ we have made explicit some
sub-asymptotic terms vanishing in the limit $r\rightarrow 0$ in such a way
that the expansion can be trusted even for $m_H$ close to $m_t$. For
practical purposes, the combined use of eq.s (\ref{9},\ \ref{10})
allows to
control numerically the $m_t^4$ terms for all values of $m_H$.
The effects of these corrections, in view of the attainable experimental
precision, start being significant for $m_t \,\,{}^\sim_>  \,\,
200 \, \mbox{GeV}$. These
corrections are being implemented in the ZFITTER code\,\,\cite{6}.

A detailed description of the calculation will be given elsewhere.

\vskip 1truecm
{\it Acknoledgements}: Useful conversations with D. Bardin and L. Maiani are
gratefully acknowledged.
\newpage
\begin{figure}[htb]
%
% HERE FIGURE BEGINS
%
% \input FEYNMAN
% \large
% \begin{picture}(5000,5000)
% \THICKLINES
% \drawline\scalar[\S\REG](14000,0)[3]
% \global\advance\pmidx by 1000
% \put(\pmidx,\pmidy){$\chi$}
% \global\advance\pmidx by -1000
% \drawline\fermion[\SE\REG](\scalarbackx,\scalarbacky)[7350]
% \global\advance\pmidx by 200
% \drawarrow[\SE\ATTIP](\pmidx,\pmidy)
% \global\advance\pmidx by 1300
% \put(\pmidx,\pmidy){$t$}
% \global\advance\pmidx by -1500
% \drawline\fermion[\SW\REG](\scalarbackx,\scalarbacky)[7350]
% \global\advance\pmidx by 200
% \drawarrow[\NE\ATTIP](\pmidx,\pmidy)
% \global\advance\pmidx by -1700
% \put(\pmidx,\pmidy){$t$}
% \global\advance\pmidx by 1500
% \drawline\scalar[\E\REG](\pbackx,\pbacky)[5]
% \drawarrow[\W\ATTIP](\pmidx,\pmidy)
% \global\advance\pmidy by -2000
% \put(\pmidx,\pmidy){$\varphi^+$}
% \global\advance\pmidy by +2000
% \drawline\fermion[\SE\REG](\pbackx,\pbacky)[5000]
% \global\advance\pmidx by 200
% \drawarrow[\LDIR\ATTIP](\pmidx,\pmidy)
% \global\advance\pmidx by 1300
% \put(\pmidx,\pmidy){$b$}
% \global\advance\pmidx by -1500
% \drawline\fermion[\SW\REG](\scalarfrontx,\scalarfronty)[5000]
% \global\advance\pmidx by 200
% \drawarrow[\NE\ATTIP](\pmidx,\pmidy)
% \global\advance\pmidx by -1700
% \put(\pmidx,\pmidy){$b$}
% \global\advance\pmidx by 1500
% \end{picture}
\vskip8truecm

% \begin{picture}(5000,5000)
% \THICKLINES
% \drawline\scalar[\E\REG](0,0)[3]
% \put(\pmidx,800){$\chi$}
% \global\advance\pbackx by 2000
% \put(\pbackx,\pbacky){\circle{4000}}
% \global\advance\pbacky by 2000
% \drawarrow[\E\ATBASE](\pbackx,\pbacky)
% \put(\pbackx,2800){$t$}
% \global\advance\pbacky by -4000
% \drawarrow[\W\ATBASE](\pbackx,\pbacky)
% \put(\pbackx,-1200){$t$}
% \global\advance\pbacky by 2000
% \global\advance\pbackx by 2000
% \drawline\scalar[\E\REG](\pbackx,\pbacky)[3]
% \put(\pmidx,800){$\chi$}
% %
% \global\advance\pbackx by 4000
% \drawline\scalar[\E\REG](\pbackx,\pbacky)[3]
% \drawarrow[\LDIR\ATBASE](\pmidx,\pmidy)
% \put(\pmidx,800){$\varphi^+$}
% \global\advance\pbackx by 2000
% \put(\pbackx,\pbacky){\circle{4000}}
% \global\advance\pbacky by 2000
% \drawarrow[\E\ATBASE](\pbackx,\pbacky)
% \put(\pbackx,2800){$t$}
% \global\advance\pbacky by -4000
% \drawarrow[\W\ATTIP](\pbackx,\pbacky)
% \put(\pbackx,-1200){$b$}
% \global\advance\pbacky by 2000
% \global\advance\pbackx by 2000
% \drawline\scalar[\E\REG](\pbackx,\pbacky)[3]
% \drawarrow[\LDIR\ATBASE](\pmidx,\pmidy)
% \put(\pmidx,800){$\varphi^-$}
% \end{picture}
%
% HERE FIGURE ENDS
%
\vskip5 truecm
\caption{relevant one loop diagrams contributing to
the constants $Z_2^\chi$, $Z_2^\varphi$ and $Z_1^\chi$}
\end{figure}
\vfill\eject

\end{document}